\documentclass[aps,prl,twocolumn,floats, floatfix]{revtex4}

\usepackage{ifthen}
\usepackage{ifpdf}

\usepackage{latexsym}
\usepackage{amsmath} 
\usepackage{amssymb} 
\usepackage{bm}
\usepackage{wasysym}

\ifpdf
\usepackage{graphicx}
\usepackage{epstopdf}
\else
\usepackage{graphicx}
\usepackage{epsfig}
\fi

\newcommand{\tbox}[1]{\mbox{\tiny #1}}

\newcommand{\be}[1]{\begin{eqnarray}\ifthenelse{#1=-1}{\nonumber}{\ifthenelse{#1=0}{}{\label{e#1}}}}
\newcommand{\ee}{\end{eqnarray}} 
\newcommand{\beq}{\begin{eqnarray}}
\newcommand{\eeq}{\end{eqnarray}} 

\newcommand{\hide}[1]{}

\begin{document}

\title{Occupation Statistics of a BEC for a Driven Landau-Zener Crossing}

\author{
Katrina Smith-Mannschott$^{1,2}$, Maya Chuchem$^{3}$, Moritz Hiller$^{4}$, Tsampikos Kottos$^{1}$ and Doron Cohen$^{3}$
}

\affiliation{
$^1$Department of Physics, Wesleyan University, Middletown, Connecticut 06459, USA \\
$^2$MPI for Dynamics and Self-Organization, Bunsenstra\ss e 10, D-37073 G\"ottingen, Germany\\
$^3$Department of Physics, Ben-Gurion University, Beer-Sheva 84105, Israel\\
$^4$\mbox{Physikalisches Institut, Albert-Ludwigs-Universit\"at, Hermann-Herder-Str.~3, D-79104 Freiburg, Germany}
}

\begin{abstract}
We consider an atomic Bose-Einstein condensate (BEC) loaded in a biased double-well trap with tunneling 
rate $K$ and interatomic interaction $U$. The BEC is prepared such that all $N$ atoms 
are in the left well. We drive the system by sweeping the potential difference~$\mathcal{E}$ between the 
two wells. 
Depending on the interaction $u=NU/K$ and the sweep rate $\dot{\mathcal{E}}$, we distinguish three dynamical regimes:
{\em adiabatic}, {\em diabatic}, and {\em sudden} and consider the occupation statistics of the final state. 
The analysis goes beyond mean-field theory and is complemented by a semiclassical picture.  
\end{abstract}
\maketitle


The theoretical and experimental study of driven atomic Bose-Einstein Condensates (BEC) in a few site 
system using optical lattice technology has intensified in recent years \cite{AGFHGO05,CSMHPR05,FTCFSWMB07,
CTFSMFB08}. Beyond the fundamental interest of these studies, they also aim to create a new generation of 
nanoscale devices such as atom transistors \cite{SAZ07}. 
Consequently, a wealth of research has been done on the prototype two-site (dimer) system, 
either within the framework of a nonlinear mean-field approach \cite{AGFHGO05}, optionally using higher order 
cumulants \cite{AV01}, or adopting a conventional many-body perspective \cite{KBK03,TWK08,HKO06,GKN08}. Such 
investigations have revealed many interesting phenomena associated with eigenvalue spectra, the structure of the 
eigenstates, wave-packet dynamics, e.g. of the Bloch-Josephson type, and leaking dynamics due to dissipative edges.

Driven dimers prove to be even more challenging \cite{WN00,LFOCN02,WGK06,SRM08,AG08}. The scenario investigated 
in these studies involves many-body Landau-Zener (LZ) transitions induced by sweeping 
the potential difference~$\mathcal{E}$ between the two wells. 
Specifically, one assumes that initially $\mathcal{E}$ is very negative 
and that all $n=N$~atoms are in the first well. 
Then $\mathcal{E}$ is increased at some constant rate $\dot{\mathcal{E}}$ 
to a very positive value.
The objective is to calculate how many atoms ($n$) remain in the first well. 
The majority of published works are based on the Gross-Pitaevskii equation (or its discrete analogue), which is a mean-field 
approach \cite{WN00,LFOCN02}. There are only a few studies that have made further progress within the framework 
of a full quantum mechanical treatment of the system \cite{WGK06,SRM08,AG08}. However, they all focus on calculating 
the average occupation $\langle n \rangle$, neglecting to form a theory for the occupation statistics $P(n)$ and, 
in particular, for the variance $\mbox{Var}(n)$.

{\bf Outline. --} 
In this Letter, we consider a driven dimer with intersite hopping amplitude $K$ and interatomic interaction $U$. 
The parameter $u=NU/K$ can be either positive (repulsive) or negative (attractive) and its magnitude
determines various dynamical regimes \cite{itemB4}: 
Rabi (${|u|<1}$), Josephson (${1<|u|<N^2}$), or Fock (${|u|>N^2}$). 
Depending on the sweep rate $\dot{\mathcal{E}}$ we distinguish 
between {\em adiabatic}, {\em diabatic}, and {\em sudden} 
dynamical scenarios and study the asymptotic occupation statistics as a function of~$u$.
Our analysis goes beyond mean-field theory, using a semiclassical picture and involving detailed simulations.

{\bf Modeling. --}
The simplest model that describes interacting bosons on a lattice 
is the Bose-Hubbard Hamiltonian (BHH), which in case of the dimer
(two sites) reads:
\begin{equation}
\mathcal{H} \!=\! 
\sum_{i=1}^{2}\!\left[\mathcal{E}_i {\hat n}_i
+\frac{U}{2}{\hat n}_i ({\hat n}_i-1)\right] \!
-\!\frac{K}{2}\sum_{i\neq j}\hat{b}_{i}^{\dagger}\hat{b}_{j}
\label{H1mode}
\end{equation}
with $\hbar{=}1$, where $\hat{b}_i$ and $\hat{b}_i^{\dagger}$ 
are bosonic annihilation and creation operators 
and ${\hat n}_i=\hat{b}_i^{\dagger}\hat{b}_i$ counts the number of particles at site~${i=1,2}$. 
The validity of this two-mode approximation \cite{itemA8a} 
for a double well is discussed in Refs.~\cite{itemA8b, itemA8c}, 
as well as in Ref.~\cite{itemC1b} for biased systems, 
and it was found to yield good agreement with the experiment \cite{itemB4}.
The total number of particles $N=\hat{n}_1+\hat{n}_2$ is a constant of motion, allowing us to 
consider a Hilbert-space of dimension ${{\cal N} =N{+}1}$, which is spanned by the Fock basis 
states ${|n\rangle\equiv|n_1{=}n,n_2{=}N{-}n\rangle}$. 
Below we assume an even $N\gg 1$ and define ${j=N/2}$, 
hence ${{\cal N} = 2j{+}1}$. The BHH for a given $N$ is formally equivalent 
to the Hamiltonian of a spin $j$~particle. 
Defining ${J_z\equiv ({\hat n}_1-{\hat n}_2)/2}$ and ${J_{+}\equiv \hat{b}_1^{\dagger} \hat{b}_2}$ 
it can be rewritten as: 
\begin{equation}
\hat{H}= U{\hat J}_z^2 +\mathcal{E} {\hat J}_z -K {\hat J}_x;\quad \mathcal{E} = \mathcal{E}_1-\mathcal{E}_2
\label{H2mode}
\end{equation}
where $\mathcal{E}$ is the bias. In the absence of interaction, this Hamiltonian can be
reinterpreted as describing  a spin in a magnetic field with precession frequency ${\Omega = (-K,0,\mathcal{E})}$.

The many-body Landau-Zener scenario assumes 
that initially all particles are located in the first site ${|\Psi(t{=}0)\rangle =|N\rangle}$.
The bias $\mathcal{E}$ is then varied with some constant sweep rate $\dot{\mathcal{E}}$. 
At the end of the sweep the occupation statistics $P(n) \equiv |\langle n|\Psi(t)\rangle|^2$
becomes time-independent. Depending on the outcome (Fig.~1) we distinguish between:
(i) an adiabatic process ${P(n)\approx\delta_{n,0}}$; 
(ii) a sudden process ${P(n)\approx\delta_{n,N}}$; 
and (iii) a diabatic process ${P(n)\approx\delta_{n,n_c}}$, 
where ${n_c \ne 0,N}$.

{\bf Phase Space. --}
In order to analyze the dynamics for finite $U$, it is convenient to rewrite the BHH using canonical variables.
Formally, our system corresponds to two coupled oscillators and thus we can define action-angle 
variables ${\hat b}_i\equiv \sqrt{{\hat n}_i} \exp(i\varphi_i)$. 
Note that the translation-like operator $\exp(i\varphi)$ is
in fact non-unitary because it annihilates the ground state, 
but this is irrelevant for ${N\gg1}$ \cite{itemB4,itemB3}.  
With these coordinates the BHH takes a form that resembles 
the Josephson Hamiltonian: 
\beq
\mathcal{H}  
\ \approx \   
\frac{NK}{2}\left[\frac{1}{2} u (\cos\theta)^2 
+ \varepsilon \cos\theta  - \sin\theta\cos\varphi \right],
\label{H3mode}
\eeq
where $\theta$ is an alternative way to express the occupation 
difference ${J_z \equiv (\hat{n}_1{-}\hat{n}_2)/2 \equiv (N/2) \cos(\theta)}$
and ${\varphi\equiv\varphi_1{-}\varphi_2}$. 
The scaled bias is $\varepsilon \equiv \mathcal{E}/K$.
Note that $\varphi$ and $\theta$ do not commute. The classical phase space is 
described either using the canonical coordinates ${(\varphi,\mathsf{n})}$, 
with ${\mathsf{n} \equiv J_z{+}(N/2) \in [0,N]}$, 
or equivalently using the spherical coordinates ${(\varphi,\theta)}$. 
In the former case the total area of phase space is~$2 \pi N$ 
with a Planck cell ${2\pi\hbar}$ and $\hbar{=}1$, while in the 
latter case the phase space has total area $4\pi$ with a Planck cell ${4\pi/N}$. 
Within the semiclassical approximation, a quantum state is described as a distribution in phase space and the eigenstates are 
associated with stripes that are stretched along contour 
lines ${\mathcal{H}(\varphi,\theta)=E}$. The 
energy levels $E_n$ can be determined via WKB quantization 
of the enclosed phase space area.

\begin{figure}[t]
\includegraphics[width=0.38\columnwidth,keepaspectratio,clip]{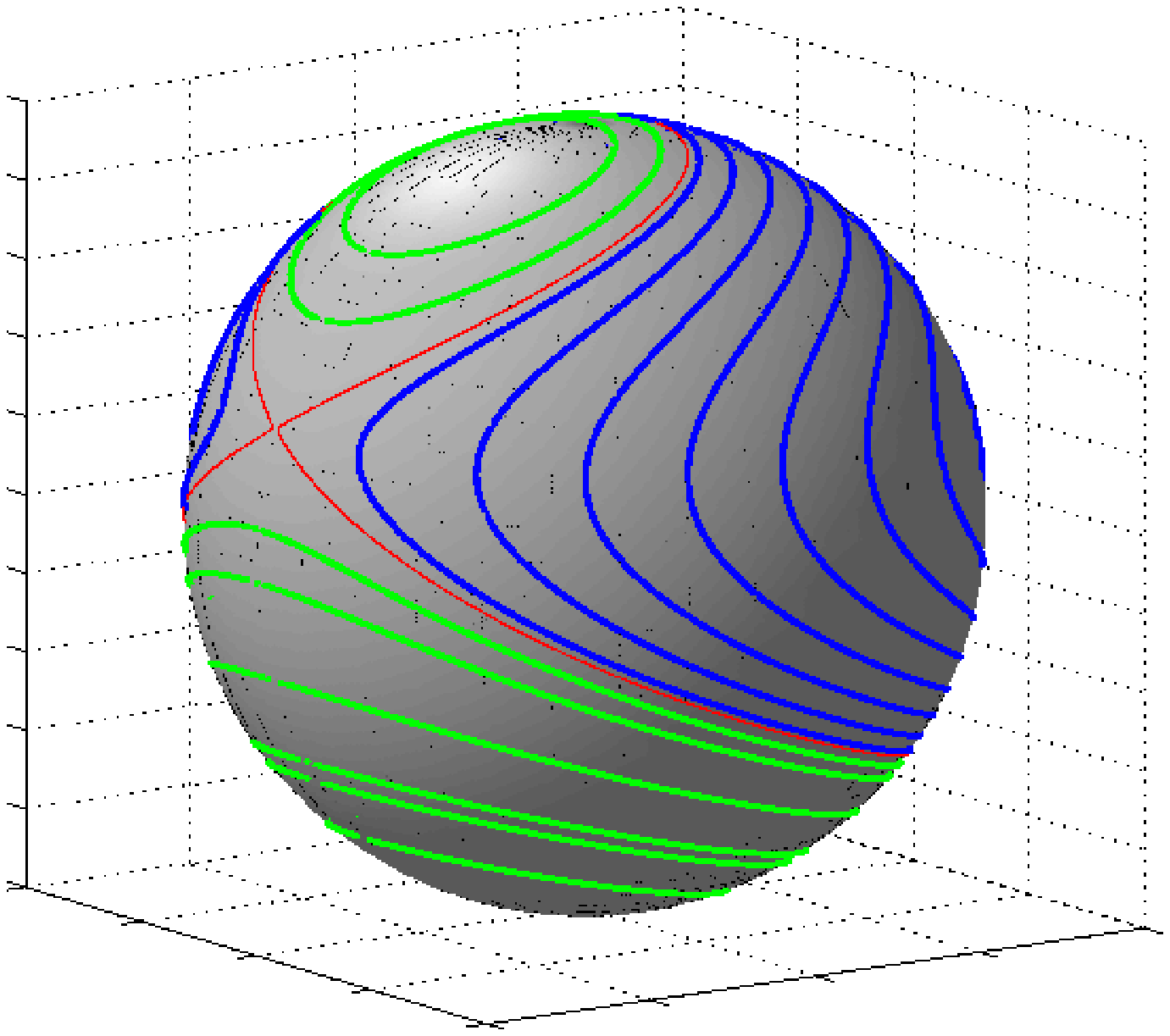} 
\includegraphics[width=0.6\columnwidth,keepaspectratio,clip]{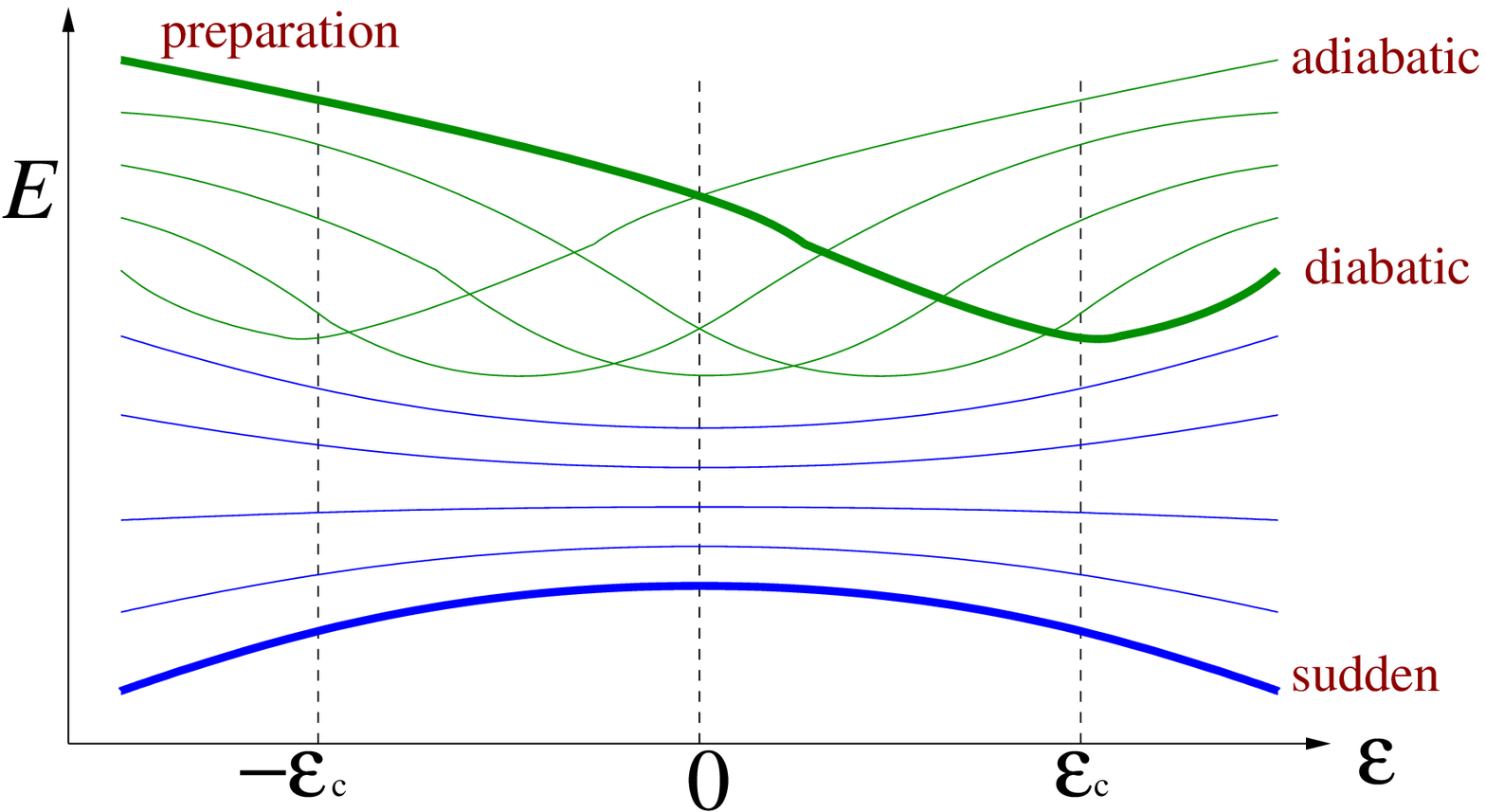} 
\caption{\label{fig1}
(Color online) 
Phase space and corresponding energy levels (conceptual plot).
Note that a ground state preparation of a ${U<0}$ system 
with increasing bias~$\mathcal{E}(t)$ is equivalent to a preparation 
in the most excited state of a $U>0$ system with decreasing bias. 
The latter convention has been adopted in the right panel and in our simulations. 
} 
\end{figure}

{\bf Separatrix. --}
The phase space topology becomes non-trivial, i.e. has more than one component as in Fig.~1a, 
if ${|u|>1}$ and $|\varepsilon|<\varepsilon_c$ where ${\varepsilon_c=(u^{2/3}-1)^{3/2}}$. 
In this regime, a separatrix divides the phase space into three regions: 
two {\em islands} (green) that contain the upper energy levels and a {\em sea} (blue) 
that contains the lower energy levels. 
In this description and in the numerics below we assume $U>0$ and adopt the following enumeration
convention: we define $E_0$ as the most excited level, while $E_N$ corresponds to the lowest one.
Note that the replacement $U\mapsto-U$ would merely invert the order of the levels, 
and $E_0$ would become the ground state.

Fig.~1b is an illustration of the energy levels as a function of the bias. Each level is associated via 
WKB quantization with one of the contour lines in Fig.~1a. 
For the sake of our later analysis we define $E_{n_c}$ as the level 
which is closest to the separatrix for $|\varepsilon|=\varepsilon_c$. 
At this critical value of the bias one of the islands has a vanishing phase space area, 
while the area of the other is ${A_c\approx 4\pi \varepsilon_c/u}$. 
Using WKB quantization we get
\beq
\label{nf}
n_c  \ \ = \ \ \frac{A_c}{4\pi/N}  \ \ \approx \ \ (1-u^{-2/3})^{3/2} \ N,
\eeq 
where the approximation for $A_c$ has been 
derived using methods as in Ref.\cite{LFOCN02}, 
and has been tested numerically.

\begin{figure}[t]
(a) \hspace*{0.24\columnwidth} (b) \hspace*{0.3\columnwidth} (c)  \\
\includegraphics[width=0.25\columnwidth,keepaspectratio,clip]{u4} 
\includegraphics[width=0.42\columnwidth,keepaspectratio,clip]{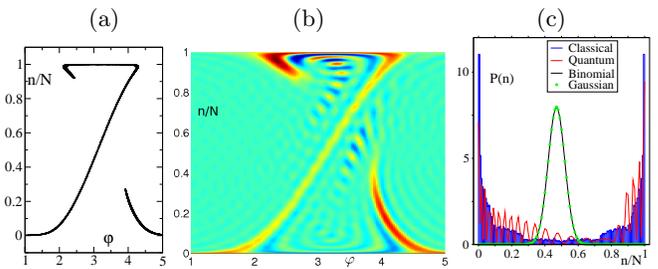} 
\includegraphics[width=0.30\columnwidth,keepaspectratio,clip]{hist_u4t5}
\caption{\label{fig2}
(Color online) Dynamical evolution in a situation 
in which the mean-field (Gaussian) approximation does not hold: 
$N{=}100$ particles are prepared in one site 
of a symmetric ($\mathcal{E}{=}0$) double well 
and then evolved with a $u{=}2$ Hamiltonian.
The initial wave-packet is stretched along the separatrix. 
(a)~(semi)classical distribution in phase space.
(b)~Wigner function of the corresponding quantum state. 
(c)~Quantum $P(n)$ compared with the (semi)classical  
and with the mean-field (binomial/Gaussian) predictions.
The average occupation $\langle n\rangle{=}47$ is associated here 
with a huge super-binomial variance $\mbox{Var}(n){=}1530$ 
instead of the binomial value $\mbox{Var}(n){\sim}25$. 
In all the simulations we use a 4th order Runge-Kutta  
and made sure that during the integration the probability 
leakage is $\ll 10^{-6}$.
} 
\end{figure}

{\bf Simulations. --}
As a preliminary step we did simulations of the undriven wave-packet 
dynamics with $\mathcal{E}{=}0$ and $u{=}2$. For this value of~$u$ 
the separatrix crosses the north pole of the phase space 
and therefore the wave-packet stretches along it, 
as illustrated in Fig.~2a (classical) and in Fig.~2b (quantum mechanical).
We note that this type of dynamics cannot be properly addressed by the
mean-field approximation. The mean-field equation merely describes the Hamiltonian evolution of a single {\em point}
in phase space and therefore assumes that the wave-packet looks like a minimal Gaussian at any moment. Whenever
the motion takes place near the separatrix, the mean-field description becomes inapplicable and 
consequently the distribution $P(n)$ is likely {\em not} to be binomial (Fig.~2c). 
In what follows we use the terms {\em sub/super-binomial} 
in order to refer to a $P(n)$ with a smaller/larger spreading than the mean-field results.
For the dynamics described in Fig.~2, the stretching along     
the separatrix leads to a super-binomial result.

Next, we address the effect of separatrix motion on $P(n)$ in the bias-sweep scenario. Note that
this separatrix motion cannot be avoided: For ${\varepsilon < -\varepsilon_c}$ the wave-packet is 
localized in the upper level. When ${\varepsilon=-\varepsilon_c}$
the separatrix emerges. As long as ${-\varepsilon_c < \varepsilon < 0}$ the wave-packet remains trapped in the top of the big
island which gradually shrinks. When $\varepsilon$ becomes larger than zero, the wave-packet can partially tunnel out
from the shrinking island to the levels of the expanding island. When ${\varepsilon=+\varepsilon_c}$, the shrinking island
disappears and the remaining part of the wave-packet is squeezed out along the $n=n_c$ contour, 
resembling the dynamics of Fig.~2.    
One observes that the stretching along the separatrix during the nonlinear LZ transition 
is accompanied by narrowing in the transverse direction. This leads to a sub-binomial rather 
than super-binomial result for the distribution $P(n)$ at the end of the sweep.

In Fig.~3 we plot the average occupation $\langle n\rangle$ 
and the participation number PN$\equiv [{\sum_n P(n)^2}]^{-1}$ of the distribution 
at the end of the sweep as a function of $\dot{\varepsilon}$. 
These, unlike $\langle n\rangle$, provides significantly more information 
regarding the nature of the crossing process. 
For very slow rates, the wave-packet follows a strict adiabatic process ending in $n{=}0$,
i.e. all particles move to the other site. For a moderate sweep rate the wave-packet ends
in a superposition of $n{=}0$ and $n{=}1$ states, indicated by PN${=}2$. 
We also resolve the possibility of ending entirely at $n{=}1$ or at $n{=}2$ or at $n{=}3$. 
In the case shown in Fig.~3, we have ${n_c\approx4}$. 
For larger sweep rates, we observe a qualitatively different behavior that
can be described as a crossover from an adiabatic/diabatic behavior to a sudden behavior 
at the peak value PN${=}4$. In order to appreciate the deviation of the numerical results 
from the mean-field theory prediction, we plot $\mbox{Var}(n)$ versus $\langle n\rangle$ 
in Fig.~4 and compare with the binomial expectation. We further analyze the observed results in the
last section.

\begin{figure}[t]
\includegraphics[width=0.85\columnwidth,keepaspectratio,clip]{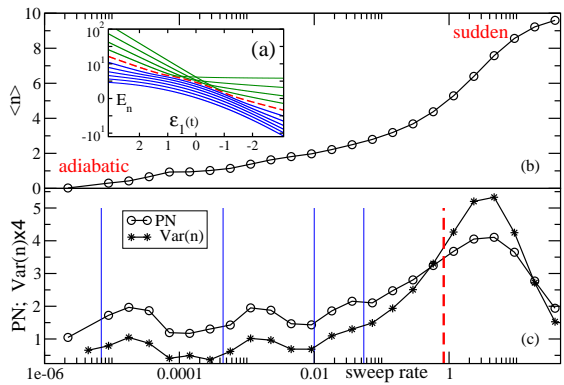} 
\caption{\label{fig3} 
(Color online) 
(a) Parametric evolution of the adiabatic energy levels 
versus the decreasing $\mathcal{E}_1(t)$ for $N{=}10$ particles and $u{=}2.5$.   
We use units of time such that $K{=}1$. 
The dashed line corresponds to the level $n_c{=}4$. 
(b) Average occupation $\langle n\rangle$ versus the sweep rate $\dot{\mathcal{E}}$.
(c) Participation number (PN) and $\mbox{Var}(n)$ versus $\dot{\mathcal{E}}$. 
The vertical lines indicate the various adiabatic and diabatic (dashed) thresholds.   
} 
\end{figure}

\begin{figure}[t]
\centerline{\includegraphics[width=0.7\columnwidth,keepaspectratio,clip]{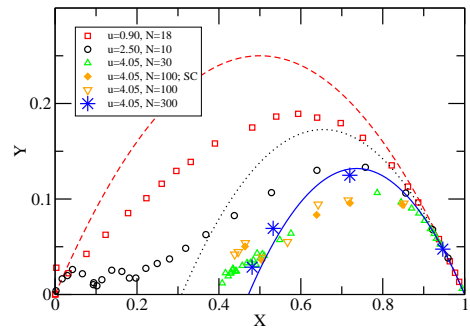}}
\caption{\label{fig4}
(Color online) $Y=\mbox{Var}(n)/N$ versus $X=\langle n\rangle/N$. 
Symbols correspond to numerical data, 
while lines indicate the sub-binomial scaling relation Eq.(\ref{scaling}) 
with $u{=}2.5$ (dotted) and $u{=}4.05$ (solid). 
No fitting is involved. 
As a reference we plot the standard binomial scaling (dashed line) 
which would strictly apply in the absence of interaction ($u{=}0$),
and the results of semiclassical (SC) simulations which are 
obtained by running ensemble of trajectories in phase space. 
There is a clear crossover from a binomial to the sub-binomial 
scaling and the agreement becomes better for large~$N$. 
Note that the super-binomial data point ${(X,Y)=(0.47,15.3)}$ 
that corresponds to the distribution in Fig.~2 is out-of-scale. 
} 
\end{figure}

{\bf Thresholds. --}
The various thresholds that are involved 
in the adiabatic-diabatic-sudden crossovers 
are indicated in Fig.~5. They all follow  
from the breakdown of a ``slowness condition" 
that can be written as 
\beq
\label{qadiab}
\dot{\mathcal{E}}  \ \ \ll \ \ \omega_{\tbox{osc}}^2  \ / \ \kappa,
\eeq
where $\omega_{\tbox{osc}}$ is a characteristic frequency of the unperturbed dynamics and $\kappa$ is the coupling 
parameter that determines the rate of the driven transitions.

In the strict quantum adiabatic framework, $\omega_{\tbox{osc}}$ 
is simply the level spacing and $\kappa$ is determined by the slopes of the intersecting levels. 
In order to determine the adiabatic thresholds in Fig.~3 
we observe that for the intersection of the $0$th level 
with the $(N{-}n)$ level the difference in slope is $\kappa=(N{-}n)$, 
because asymptotically ${d(E_n{-}E_m)/d\mathcal{E} \sim (n{-}m)}$. 
In the absence of interaction ($u \ll 1$), the
level spacing is $\omega_{\tbox{osc}}=K$ and only nearby levels are coupled, leading to the standard Landau-Zener adiabaticity
condition ${\dot{\mathcal{E}} \ll K^2}$. With strong interaction there is an $N$th order coupling 
between the $n{=}0$ level and the $n{=}N$ level, 
which allows tunneling from the top of one island to the top of the 
other island (as illustrated in Fig.~1). An estimate for this coupling 
is $K_{\tbox{eff}}=[NK]/[2^{N{-}1} (N{-}1)!](K/U)^{N{-}1}$ \cite{KBK03}. 
Similar considerations can be applied to the bottom sea level leading to 
the distinction between mega, gradual, and sequential crossings (see Ref.\cite{HKC08}).

For large $N$ it might be practically impossible to satisfy the strict adiabatic condition 
which is associated with the possibility to tunnel from the top of one island to the top of the other. 
Then the relevant mechanism for transition, i.e. the emission to the level $n_c$ as described 
in the previous section, becomes semiclassical.
The frequency that governs this process is the oscillation frequency 
at the bottom of the sea ${\omega_{\rm osc} \sim |NUK|^{1/2}}$. 
It determines the level spacing of the lower energy levels and also describes the level spacing in 
the vicinity of the separatrix, apart from some logarithmic corrections \cite{csp}. 
It follows that the diabatic-sudden crossover involves the threshold condition ${\dot{\mathcal{E}} \ll |NUK|}$ 
as indicated in Fig.~5 and in Fig.~3 for the specific parameters of the simulations.

\begin{figure}[t]
\includegraphics[width=\columnwidth,keepaspectratio,clip]{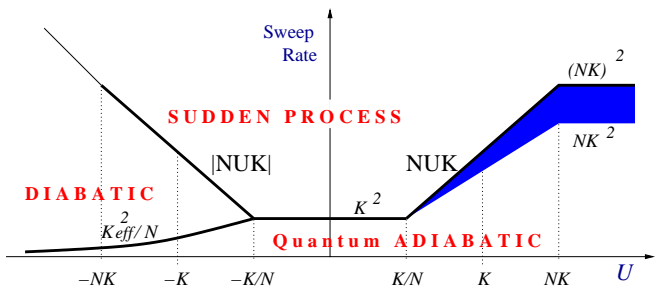} 
\caption{\label{fig5}
(Color online) 
Diagram of the $(U,\dot{\mathcal{E}})$ regimes 
for a ground state preparation. 
In the Rabi regime $|U|<K/N$ we have a
crossover from adiabatic to sudden behavior. 
For ${K/N<U<NK}$ (or ${U>NK}$) 
we have a broad crossover from adiabatic
gradual (or sequential \cite{HKC08}) behavior 
to sudden behavior. For ${U<-K/N}$ we have two crossovers: 
the first from quantum adiabatic to diabatic behavior 
and the second from diabatic to sudden behavior. 
The diabatic behavior can be regarded as a classical nonlinear 
adiabatic behavior. For ${U<-NK}$ the distinction between 
the diabatic and the sudden regime is blurred
because the final state is the same.
} 
\end{figure}

{\bf Scaling. --}
Due to the squeezing along the separatrix, the spreading of the wavepacket 
for an idealized diabatic process becomes negligible in the transverse direction. 
The diabatic-sudden crossover is related to the non-adiabatic 
transitions between the remaining $(N{-}n_c)$ {\em sea} levels, 
where nonlinear effects are negligible. It follows that the spreading   
can be approximately modeled by the toy Hamiltonian $\mathcal{H}= h(t) \cdot J$, 
where $J$ is a spin entity with $j_{\tbox{eff}}=(N{-}n_c)/2$, 
and $h(t)$ is a field with constant magnitude $|h(t)|=\omega_{\tbox{osc}}$ 
corresponding to the mean-level spacing. 
The sweep is like a rotation of $h(t)$ in the plane with some
angular rate $\omega$. For such a (linear) model the mean-field approximation 
is exact and therefore we suggest (due to the truncation of Hilbert space)  
a sub-binomial rather than binomial scaling relation between the mean and the
variance of the occupation statistics: 
\beq
\label{scaling}
Y \ = \ (1-X) \frac{X-c}{1-c} ; \quad {\rm with} \quad c=\frac{n_c}{N},
\eeq
where ${X=\langle n \rangle/N}$ and ${Y=\mbox{Var}(N)/N}$. Our numerical data is reported in Fig.~4 together with the binomial
($c{=}0$) and sub-binomial scaling relation Eq.(\ref{scaling}). The numerics confirm the expected $u$-dependent crossover from
binomial to sub-binomial statistics, where the latter, with no fitting parameters, sets a lower bound for the variance.

{\bf Summary. --}
In view of the strong research interest in counting statistics of electrons in mesoscopic devices, it is surprising that 
the issue of occupation statistics of BECs has been explored only for equilibrium phase transitions. We were motivated to
address this subject in the framework of dynamical processes by state-of-the-art experiments aimed at counting individual 
particles \cite{CSMHPR05,FTCFSWMB07,CTFSMFB08}. We have shown that in the case of a many-body Landau-Zener transition the 
mean-field binomial expectation is not realized, however a sub-binomial scaling relation still works quite well. The study 
of the occupation statistics and, in particular, the participation number of the final distribution, allowed us to resolve 
all the details of the adiabatic-diabatic-sudden crossovers and to verify theoretical estimates for the threshold of each 
crossover.

{\bf Acknowledgments. --}
This research was funded by a grant from the US-Israel Binational Science Foundation (BSF), 
the DFG Forschergruppe 760, and a DIP grant.

\vspace*{-3mm}



\clearpage


\begin{thebibliography}{99}

\vspace*{-4mm}

\bibitem{AGFHGO05} 
M. Albiez, et al., Phys. Rev. Lett. 95, 010402 (2005).

\bibitem{CSMHPR05}
C.-S. Chuu, F. Schreck, T.P. Meyrath, J.L. Hanssen, G.N. Price, and M.G. Raizen,
Phys.  Rev. Lett. 95, 260403 (2005);
A.M. Dudarev, M.G. Raizen, and Q. Niu, Phys. Rev. Lett. 98, 063001 (2007).

\bibitem{FTCFSWMB07}
S. Foelling, S. Trotzky, P. Cheinet, M. Feld, R. Saers, A. Widera, T. Mueller, I. Bloch, Nature 448, 1029 (2007)

\bibitem{CTFSMFB08}
P. Cheinet, S. Trotzky, M. Feld, U. Schnorrberger, M. Moreno-Cardoner, S. Foelling, I. Bloch, arXiv:0804.3372

\bibitem{SAZ07} 
J. A. Stickney, D. Z. Anderson, A. A. Zozulya, Phys. Rev. A 75, 013608 (2007).

\bibitem{AV01}
J. R. Anglin and A. Vardi, Phys. Rev. A 64, 013605 (2001);
A. Vardi, V. A. Yurovsky, and J.R. Anglin, Phys. Rev. A 64, 063611 (2001);
A. Vardi and J. R. Anglin, Phys. Rev. Lett. 86, 568 (2001).

\bibitem{KBK03}
G. Kalosakas, A. R. Bishop, and V. M. Kenkre, Phys. Rev. A 68, 023602 (2003);
G Kalosakas, A R Bishop and V M Kenkre, J. Phys. B: At. Mol. Opt. Phys. 36, 3233 (2003);
G. Kalosakas and A. R. Bishop, Phys. Rev. A 65, 043616 (2002).

\bibitem{TWK08}
F. Trimborn, D. Witthaut, H. J. Korsch, arXiv:0802.1142.

\bibitem{HKO06}
M. Hiller, T. Kottos, and A. Ossipov, Phys. Rev. A {\bf 73}, 063625 (2006).

\bibitem{GKN08} 
E. M. Graefe, H. J. Korsch, A. E. Niederle, Phys. Rev. Lett. 101, 150408 (2008).

\bibitem{LFOCN02}
J. Liu, L.-B. Fu, B.-Y. Ou, S.-G. Chen, and Q. Niu, 
Phys. Rev. A 66, 023404 (2002).

\bibitem{WN00}
B. Wu and Q. Niu, 
Phys. Rev. A 61, 023402 (2000).

\bibitem{WGK06}
D. Witthaut, E. M. Graefe, and H. J. Korsch, Phys. Rev. A 73, 063609 (2006) .

\bibitem{SRM08} 
P. Solinas, P. Ribeiro, R. Mosseri, arXiv:0807.0703.

\bibitem{AG08}  
A. Atland, V. Gurarie, Phys. Rev. Lett. {\bf 100}, 063602 (2008).

\bibitem{itemB4}
R. Gati and M. K. Oberthaler, J. Phys. B 40 (2007) R61.

\bibitem{itemA8a}
G. J. Milburn, J. Corney, E. M. Wright, and D. F. Walls
Phys. Rev. A 55, 4318 (1997).

\bibitem{itemA8b}
R. W. Spekkens and J. E. Sipe, Phys. Rev. A 59, 3868 (1999).

\bibitem{itemA8c}
D. Ananikian and T. Bergeman, Phys. Rev. A 73, 013604 (2006).

\bibitem{itemC1b} 
D. R. Dounas-Frazer, A.M. Hermundstad, and L. D. Carr, Phys. Rev. Lett. 99, 200402 (2007)

\bibitem{itemB3} 
A. J. Leggett, Rev. Mod. Phys. 73, 307 (2001).

\bibitem{HKC08}
M. Hiller, T. Kottos and D. Cohen, Europhys. Lett. {\bf 82}, 40006 (2008); 
M. Hiller, T. Kottos and D. Cohen, Phys. Rev. A {\bf 78}, 013602 (2008).

\bibitem{csp}
E. Boukobza, M. Chuchem, D. Cohen and A. Vardi, 
Phys. Rev. Lett. 102, 180403 (2009).


\end{thebibliography}
\end{document}